\documentclass[aps,prl,superscriptaddress,reprint]{revtex4-1}
\usepackage{amsmath}
\usepackage{amssymb}
\usepackage{graphicx}
\usepackage[colorlinks,urlcolor=blue]{hyperref}
\usepackage{url}
\usepackage{color,xcolor}
\usepackage{ulem}
\usepackage{float}
\usepackage{epstopdf}
\begin{document}

\title{Iron telluride ladder compounds: \\ Predicting the structural and magnetic properties of BaFe$_2$Te$_3$}
\author{Yang Zhang}
\author{Ling-Fang Lin}
\affiliation{Department of Physics and Astronomy, University of Tennessee, Knoxville, TN 37996, USA}
\affiliation{School of Physics, Southeast University, Nanjing 211189, China}
\author{Adriana Moreo}
\affiliation{Department of Physics and Astronomy, University of Tennessee, Knoxville, TN 37996, USA}
\affiliation{Materials Science and Technology Division, Oak Ridge National Laboratory, Oak Ridge, TN 37831, USA}
\author{Shuai Dong}
\affiliation{School of Physics, Southeast University, Nanjing 211189, China}
\author{Elbio Dagotto}
\email{Corresponding author: edagotto@utk.edu}
\affiliation{Department of Physics and Astronomy, University of Tennessee, Knoxville, TN 37996, USA}
\affiliation{Materials Science and Technology Division, Oak Ridge National Laboratory, Oak Ridge, TN 37831, USA}

\date{\today}

\begin{abstract}
Since the discovery of pressure-induced superconductivity in the two-leg ladder system BaFe$_2X_3$ ($X$=S, Se), with the 3$d$ iron electronic density $n = 6$, the quasi-one-dimensional iron-based ladders have attracted considerable attention. Here, we use Density Functional Theory (DFT) to predict that the novel $n = 6$ iron ladder BaFe$_2$Te$_3$ could be stable with a similar crystal structure as BaFe$_2$Se$_3$. Our results also indicate that BaFe$_2$Te$_3$ will display the complex 2$\times$2 Block-type magnetic order. Due to the magnetic striction effects of this Block order, BaFe$_2$Te$_3$ should be a magnetic noncollinear ferrielectric system with a net polarization $0.31$ $\mu$C/cm$^2$. Compared with the S- or Se-based iron ladders, the electrons of the Te-based ladders are more localized, implying that the degree of electronic correlation is enhanced for the Te case which may induce additional interesting properties. The physical and structural similarity with BaFe$_2$Se$_3$ also suggests that BaFe$_2$Te$_3$ could become superconducting under high pressure.
\end{abstract}

\maketitle

\section{I. Introduction}
Since the initial discovery of superconductivity in fluorine-doped two-dimensional (2D) iron square lattices LaFeAsO, iron based compounds have rapidly developed into one of the most important branches of unconventional superconductors in Condensed Matter Physics and Material Science~\cite{Stewart:Rmp,Johnston:Ap,Dagotto:Rmp,Dai:Rmp}. In the nonsuperconducting parent compounds, their magnetic ground state is known as collinear stripe order, i.e. the so-called C-type antiferromagnetic (AFM) order, with some exceptions such as Block-type and Bicollinear-type~\cite{Dai:Rmp,Dagotto:Rmp,Li:Np,Zhang:RRL}. Since most superconducting phases always emerge next to the suppression of antiferromagnetism by carrier doping or pressure, the AFM order and AFM spin fluctuations are considered to be important for superconductivity~\cite{Mazin:np,Dai:Np}.

Recently, pressure-induced superconductivity was observed in the two-leg quasi-one-dimensional ladder system BaFe$_2$$X_3$ ($X$ = S, Se) with eletronic density $n=6.0$~\cite{Takahashi:Nm,Ying:prb17}, starting a novel field of research for high temperature iron-based superconductors~\cite{Yamauchi:prl15,Arita:prb,Patel:prb16,Wang:prb16,Zhang:prb17,chi:prl,Patel:prb17,Pizarro:prm,Zheng:prb18,Zhang:prb18,Zhang:prb19}. BaFe$_2$S$_3$ displays a stripe-type (CX) AFM order, similar to other 2D iron-based superconductors, below $120$ K~\cite{Takahashi:Nm,chi:prl}.
By applying hydrostatic pressure, this system displayed an insulator-metal transition~\cite{Suzuki:prb,Zhang:prb17} and superconductivity was observed at $P\sim 11$~GPa near a first-order magnetic phase transition where the AFM order was suppressed~\cite{Takahashi:Nm,Materne:prb19}. These recent developments in the context of two-leg iron ladders remind us of the previous results for copper ladders where superconducting tendencies upon doping were theoretically predicated and later confirmed experimentally~\cite{cu-ladder1,cu-ladder2,cu-ladder3}. In fact, in iron ladders similar pairing tendencies were observed upon hole doping by
theoretical research based on density matrix renormalization group (DMRG) at intermediate Hubbard coupling strengths~\cite{Patel:prb16,Patel:prb17}. Another possible explanation of superconductivity is the originally-proposed band narrowing by pressure~\cite{Takahashi:Nm}.

BaFe$_2$Se$_3$ also shows pressure-induced superconductivity~\cite{Ying:prb17,Zhang:prb18}, but in addition it displays an exotic 2$\times$2 Block-type magnetic order below $256$~K under ambient conditions~\cite{Caron:Prb12}. Due to its enhanced degree of electronic correlation~\cite{Caron:Prb,Nambu:Prb}, the physical properties of BaFe$_2$Se$_3$ become more complex. For example, the existence of an ``orbital-selective Mott phase'' (OSMP) was found by neutron experiments at ambient pressure~\cite{mourigal:prl15}. This OSMP of BaFe$_2$Se$_3$ was theoretically discussed based on DMRG methods by using multiorbital Hubbard models as well~\cite{osmp1,osmp2,osmp3}.
However, an additional striking experimental discovery was recently reported for this interesting material: the existence of a polar state with possible polar orbital ordering was confirmed by neutron diffraction methods combined with optical second harmonic generation signals~\cite{Aoyama:prb19}. In fact, this material was first theoretically predicted to be multiferroic because the Block-type magnetic order can produce displacements of Se inducing broken inversion symmetry~\cite{Dong:PRL14}. It should be noted that BaFe$_2$Se$_3$ is the first reported iron-based system to become both superconductor and multiferroic. Moreover, the polar state of BaFe$_2$Se$_3$ is an exotic noncollinear ferrieletric (FiE) phase instead of a plain ferroelectric (FE) one~\cite{Dong:PRL14,Aoyama:prb19}. To our best knowledge, noncollinear FiE order was only proposed in a few compounds, such as $M$O$_2$$X_2$ ($M$= Mo/W, $X$=Br/Cl)~\cite{Lin:PRL} and strained BiFeO$_3$~\cite{Yang:PRL}. It is reasonable to assume that finding this exotic 2$\times$2 Block-type magnetic order with quasi one-dimensional ladders defines an effective feasible path to explore FiE materials. Besides, considering the similar atomic average electronic density in many iron superconductors, it is conceivable to obtain the superconducting phase in other potential $n = 6$ ladders as well.

However, to our best knowledge, there is no other iron ladder reporting to display the 2$\times$2 Block-type magnetic order. According to our previous Hartree-Fock~\cite{luo:prb13} and DFT calculations~\cite{Zhang:prb19}, the Block-type was expected to be stable in a large region of the Hund coupling $J_H$ and Hubbard $U$ phase diagram in $n=6$ iron ladders. Hence, it can be reasonably assumed that the magnetic ground state of $n=6$ Te-based iron ladders, not synthesized  yet, may also display the Block-type order if it can be prepared. Actually, the $n=5.5$ iron Te-based ladder was synthesized in experiments~\cite{Klepp:JOAC} but it was recently predicted to display CX-type magnetic order~\cite{Zhang:prb19}. Considering that the ionic radius of Rb$^+$ ($\sim1.47$ \AA) and Ba$^{\rm 2+}$ ($\sim1.434$ \AA) are similar, we believe it should be possible to prepare Te-based $n=6$ iron ladders with chemical formula BaFe$_2$Te$_3$.

In the present publication, we performed first-principles DFT calculations for the BaFe$_2$Te$_3$ system. Our theoretical results indicate that BaFe$_2$Te$_3$ should be stable with a similar crystal structure as BaFe$_2$Se$_3$.
Because in the past DFT has successfully predicted many new compounds before they were truly prepared, such as blue phosphorene~\cite{Guan:prl} and phosphorus carbide~\cite{Guan:nl,Tan:AM}, our structural prediction should be reliable.
Moreover, the 2$\times$2 Block-type spin order is also predicted to be the most likely magnetic ground state in our calculations for this compound. In addition, we found that BaFe$_2$Te$_3$ should display noncollinear ferrielectric order driven by the 2$\times$2 Block-type magnetic order via magnetic exchange striction. The magnetic state and electronic structure similarity with BaFe$_2$Se$_3$ further suggests that BaFe$_2$Te$_3$ could also become superconducting under high pressure.

\section{II. Method}
To understand the physical properties of the BaFe$_2$Te$_3$ system, first-principles DFT calculations were performed based on the projector augmented wave (PAW) pseudopotentials with the Perdew-Burke-Ernzerhof (PBE) exchange functional, as implemented in the Vienna {\it ab initio} simulation package (VASP) code~\cite{Kresse:Prb,Kresse:Prb96,Blochl:Prb,Perdew:Prl}. Since the spin-polarized PBE-GGA function is known to provide an accurate description of the iron based $123$-type two-leg ladder systems~\cite{Suzuki:prb,Arita:prb,Zhang:prb17,Zhang:prb18,Zheng:prb18}, in this publication we do not add an additional Hubbard $U$.

Due to the quasi-one-dimensional ladder structure, the magnetic coupling in-ladder should be the dominant factor affecting the energies and physical properties. Various possible (in-ladder) magnetic configurations were imposed on the iron ladders~\cite{Zhang:prb19} [see Fig.~S1 of the Supplemental Material (SM)~\cite{Supplemental}] to predict the magnetic properties, such as nonmagnetic (NM), ferromagnetic (FM), CX-type with FM rungs and AFM legs, CY-type with AFM rungs and FM legs, G-type with both AFM rungs and legs, and 2$\times$2 magnetic Block-type. Considering previous neutron results for two-leg iron ladders~\cite{Takahashi:Nm,Caron:Prb12}, the ($\pi$, $\pi$, $0$) order was adopted for the CX-AFM and Block-AFM orders. The plane-wave cutoff energy was $500$ eV. Since different magnetic configurations have different minimal unit cells, the mesh was appropriately modified for all the candidates to render the $k$-point densities approximately the same in reciprocal space (as example, $7\times5\times11$ for the FM-type). In addition, we have tested that these $k$-point meshes already lead to converged energies when compared with denser meshes. Both the lattice constants and atomic positions were fully relaxed with different spin configurations until the Hellman-Feynman force on each atom was smaller than $0.01$ eV/{\AA}.

The phonon spectra was calculated using the finite displacement approach and analyzed by the PHONONPY software~\cite{Chaput:prb,Togo:sm}. Furthermore, to estimate the FE polarization, the Berry phase method was adopted~\cite{King-Smith:Prb,Resta:Rmp}. In addition to the standard DFT calculation discussed thus far, the maximally localized Wannier functions (MLWFs) method was employed to fit Fe $3d$'s five bands by using the WANNIER90 packages~\cite{Mostofi:cpc}.

\section{III. Results}

\subsection{A. Crystal structure}
\begin{figure}
\centering
\includegraphics[width=0.48\textwidth]{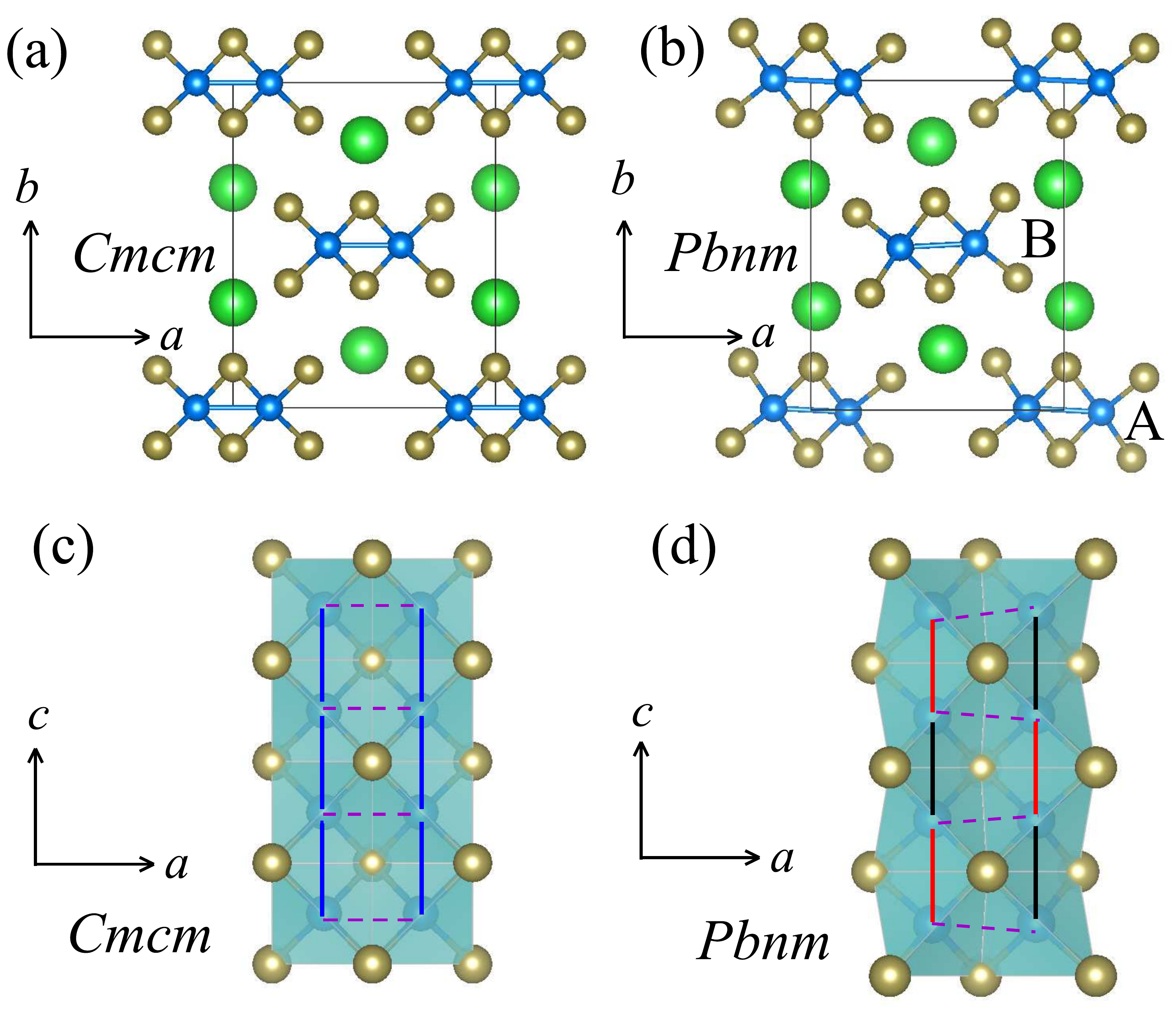}
\caption{Schematic crystal structure of BaFe$_2$Te$_3$ (electronic density $n = 6$) with the convention: Green = Ba; Blue = Fe; Dark yellow = Te. For better comparison, we used the space group $Pbnm$ instead of the conventional $Pnma$ since the lattice vectors of the $Pbnm$ space group are the same as in the $Cmcm$ space group. Note that the difference between Pbnm and Pnma space groups is only regarding the choice of a unique axis. (a-b) Sketch of the possible crystal structures of BaFe$_2$Te$_3$ for the $Cmcm$ (No. 63) and $Pbnm$ (No. 62) phases, respectively. (c-d) One iron ladder with highlighted FeTe$_4$ tetrahedra for the $Cmcm$ and $Pbnm$ phases, respectively.}
\label{Fig1}
\end{figure}

As shown in Fig.~\ref{Fig1}, we constructed two crystal structures [$Cmcm$ (No.63) and $Pbnm$ (No.62) phases] for BaFe$_2$Te$_3$, because those two types of phases were found experimentally in other $123$-type iron ladders~\cite{Caron:Prb12,Svitlyk:JPCM}. In the $Cmcm$ phase [see Fig.~\ref{Fig1}(a)], the FeTe$_4$ tetrahedra are aligned in the $ac$ plane. The $Pbnm$ phase [see Fig.~\ref{Fig1}(b)] can be visualized as adding a tilting of the FeTe$_4$ tetrahedra along the $c$-axis on the $Cmcm$ phase of BaFe$_2$Te$_3$, while the two FeTe$_4$ tetrahedra along the rung direction are rotated counterclockwise/clockwise, respectively. As a consequence, as shown in Figs.~\ref{Fig1}(c-d), the iron ladder would slightly distort with two different iron-iron distances along the leg direction in the $Pbnm$ phase while the iron-iron distances are equal in the ideal $Cmcm$ ladder. By comparing the energies of the two phases with a non-magnetic state, the $Pbnm$ phase was considered to be the most likely crystal structure of BaFe$_2$Te$_3$ ($\sim1.3$ meV/f.u. lower than $Cmcm$ phase) after the lattice constants and atomic positions were fully relaxed. However, note that the energy difference of the two phases for the non-magnetic state are quite small and this issue will be discussed in the next section.

For completeness, starting from the crystal lattice with $Cmcm$ (No.63) and $Pbnm$ (No.62) symmetry plus various magnetic states, the different spin configurations were fully relaxed. The DFT results indicate that the crystal structure of BaFe$_2$Te$_3$ should be similar to BaFe$_2$Se$_3$ instead of BaFe$_2$S$_3$. In the non-tilting ladder structure, all the energies of different magnetic configurations were higher than the corresponding energies of the tilting ladder structure [see Table~\ref{Table1} and Table~\ref{Table3}]. Hence, we conclude that the $Pbnm$ (No.62) phase should be the favored structure of BaFe$_2$Te$_3$.

\begin{table}
\centering\caption{The optimized lattice constants ({\AA}), local magnetic moments (in $\mu_{\rm B}$/Fe units) within the default PAW sphere, and band gaps (eV) for the various magnetic configurations using the $Cmcm$ structure. Also included the energy differences (meV/Fe) with respect to the Block-B AFM configuration in the $Pbnm$ phase, taken as the reference of energy.}
\begin{tabular*}{0.48\textwidth}{@{\extracolsep{\fill}}llllc}
\hline
\hline
  & $a$/$b$/$c$ & $M$ & Gap  & Energy \\
\hline
NM       & 9.738/11.991/5.672  & 0    & 0  & 385.8   \\
FM       & 9.805/13.195/5.634  & 2.92 & 0  & 155.7    \\
CX       & 9.805/12.825/5.703  & 2.60 & 0 & 84.3 \\
CY       & 9.756/13.214/5.630  & 2.77 & 0.22 & 61.8   \\
G        & 9.799/12.812/5.682  & 2.54 & 0 & 160.7    \\
Block-A  & 9.837/13.203/5.632  & 2.87 & 0.17 & 44.0   \\
Block-B  & 9.823/13.131/5.652  & 2.85 & 0.3&    24.9 \\
\hline
\hline
\end{tabular*}
\label{Table1}
\end{table}

\subsection{B. Stability}
\begin{figure}
\centering
\includegraphics[width=0.48\textwidth]{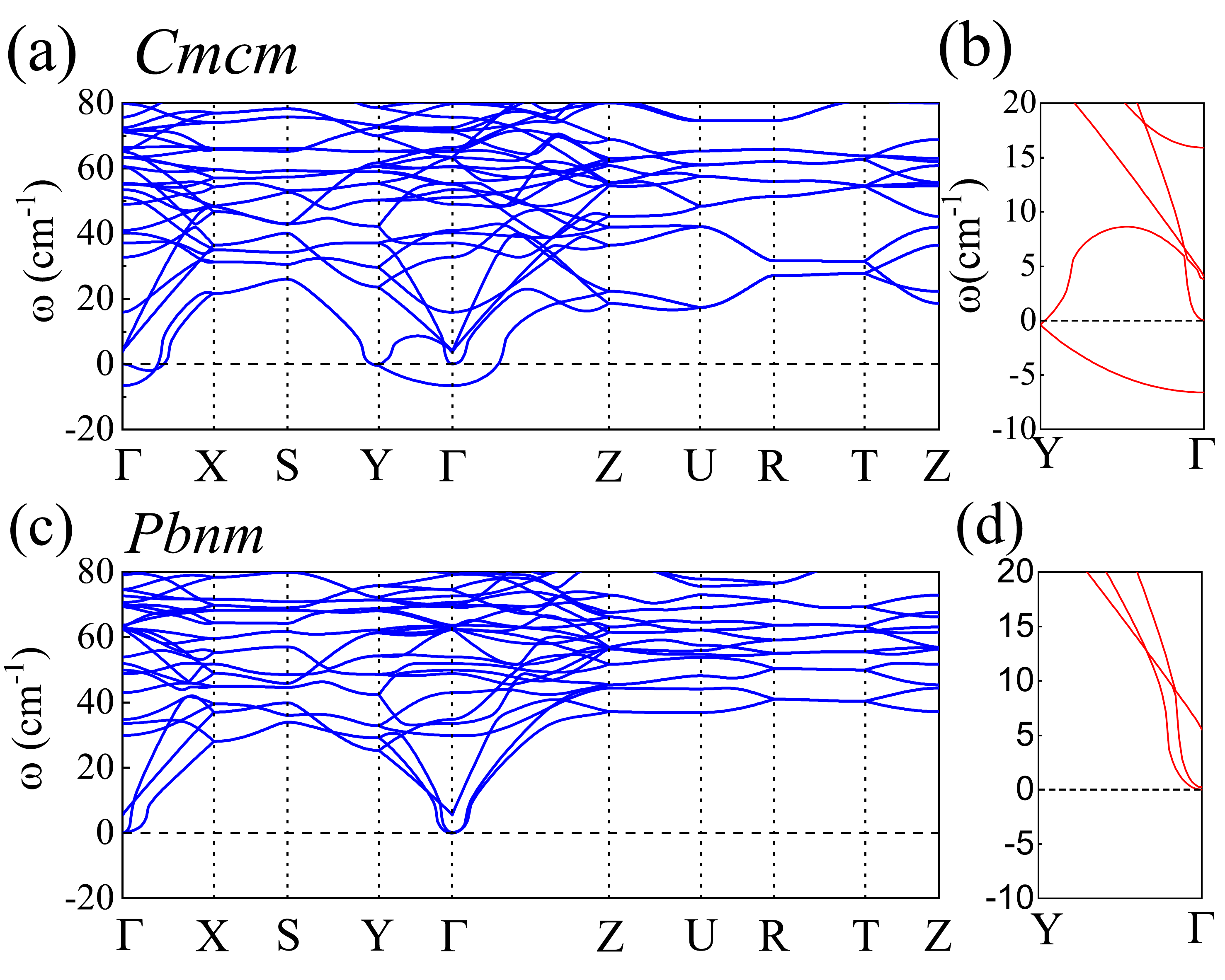}
\caption{ Phonon spectrum of BaFe$_2$Te$_3$ (electronic density $n = 6$) for NM state. The coordinates of the high symmetry points in the bulk Brillouin zone (BZ) are given by: $\Gamma$ = (0, 0, 0), X = (0.5, 0, 0), S = (0.5, 0.5, 0), Y = (0, 0.5, 0), Z = (0, 0, 0.5), U = (0, 0, 0.5), R = (0.5, 0.5, 0.5), T (0, 0.5, 0.5). (a) Sketch of the entire BZ for the $Cmcm$ phase of BaFe$_2$Te$_3$. (b) The phonon spectrum of the $Cmcm$ phase along Y-$\Gamma$. (c) Sketch of the entire BZ for the $Pbnm$ phase of BaFe$_2$Te$_3$. (d) The phonon spectrum of the $Pbnm$ phase along Y-$\Gamma$.}
\label{Fig2}
\end{figure}

As shown in Figs.~\ref{Fig2}(a-b), there is a small imaginary-frequency branch in the phonon spectrum of the $Cmcm$ structure of BaFe$_2$Te$_3$, which will lead to spontaneous distortions. By removing this unstable phonon modes, the symmetry decreases from $Cmcm$ to $Pbnm$. This small imaginary-frequency issue also corresponds to the small energy difference between the $Cmcm$ and $Pbnm$ phases. According to group theory analysis using the AMPLIMODES software~\cite{Orobengoa:jac,Perez-Mato:aca}, this spontaneous distortion mode is a Y$^{\rm 2+}$ mode. For comparison, the phonon spectrum of the $Pbnm$ structure of BaFe$_2$Te$_3$ is displayed in Figs.~\ref{Fig2}(c-d), which is dynamically stable (no unstable modes). Furthermore, we also investigated the elastic-stability conditions that indicated the $Pbnm$ structure of BaFe$_2$Te$_3$ should be elastically stable.

The elastic matrix of BaFe$_2$Te$_3$ has nine non-zero independent matrix elements ($C_{\rm 11}$, $C_{\rm 12}$, $C_{\rm 13}$, $C_{\rm 22}$, $C_{\rm 23}$, $C_{\rm 33}$, $C_{\rm 44}$, $C_{\rm 55}$, $C_{\rm 66}$) due to the $mmm$ Laue class features of the $Cmcm$ space group (No. 63) and $Pbnm$ space group (No. 62)~\cite{Mouhat:PRB}. These values of the elastic matrix constants $C_{ij}$'s satisfy the Born stability criteria for an orthorhombic system~\cite{Mouhat:PRB}. We calculated the elastic matrix for both $Cmcm$ and $Pbnm$ phase as summarized in Table~\ref{Table2}.

\begin{table} [H]
\centering\caption{The calculated elastic matrix elements ($C_{ij}$, in units of Gpa) corresponding to BaFe$_2$Te$_3$ for the $Cmcm$ and $Pbnm$ phases.}
\begin{tabular*}{0.48\textwidth}{@{\extracolsep{\fill}}lllllllllc}
\hline
\hline
  & $C_{\rm 11}$ & $C_{\rm 12}$ &$C_{\rm 13}$ & $C_{\rm 22}$ &$ C_{\rm 23}$ & $ C_{\rm 33}$ &$ C_{\rm 44}$ & $ C_{\rm 55}$ & $C_{\rm 66}$ \\
\hline
No. 63       & 48.4 & 21.1 & 30.8 & 34.3 & 28.2 & 90.8 & 15.3  & 23.5  & 12.9   \\
No. 62       & 48.8 & 23.7 & 29.1 & 38.8 & 27.4 & 88   & 15.7  & 23.1  & 13.9   \\
\hline
\hline
\end{tabular*}
\label{Table2}
\end{table}

The necessary and sufficient Born criteria for an orthorhombic system are the following~\cite{Mouhat:PRB}:

(1) The matrix $C$ is definite positive;

(2) all eigenvalues of $C$ are positive;

(3) all the leading principal minors of $C$ (determinants of its upper-left $k \times k$ submatrix, 1$\leq$$k$$\leq$6) are positive.

(4)\begin{equation}
C_{\rm 11}>0, C_{\rm 11}C_{\rm 22}>C_{\rm 12}^2, \\[-3mm]
\end{equation}
\begin{equation}
C_{\rm 11}C_{\rm 22}C_{\rm 33}+2C_{\rm 12}C_{\rm 13}C_{\rm 23}-C_{\rm 11}C_{\rm 23}^2-C_{\rm 22}C_{\rm 13}^2-C_{\rm 33}C_{\rm 12}^2>0, \\[-1mm]
\end{equation}
\begin{equation}
C_{\rm 44}>0, C_{\rm 55}>0, C_{\rm 66}>0, \\[-4mm]
\end{equation}
\begin{equation}
C_{ii}+C_{jj}-2C_{ij}>0, \\[-1mm]
\end{equation}

As a consequence, BaFe$_2$Te$_3$ should be elastically stable.

\subsection{C. Magnetism}
\begin{figure}
\centering
\includegraphics[width=0.48\textwidth]{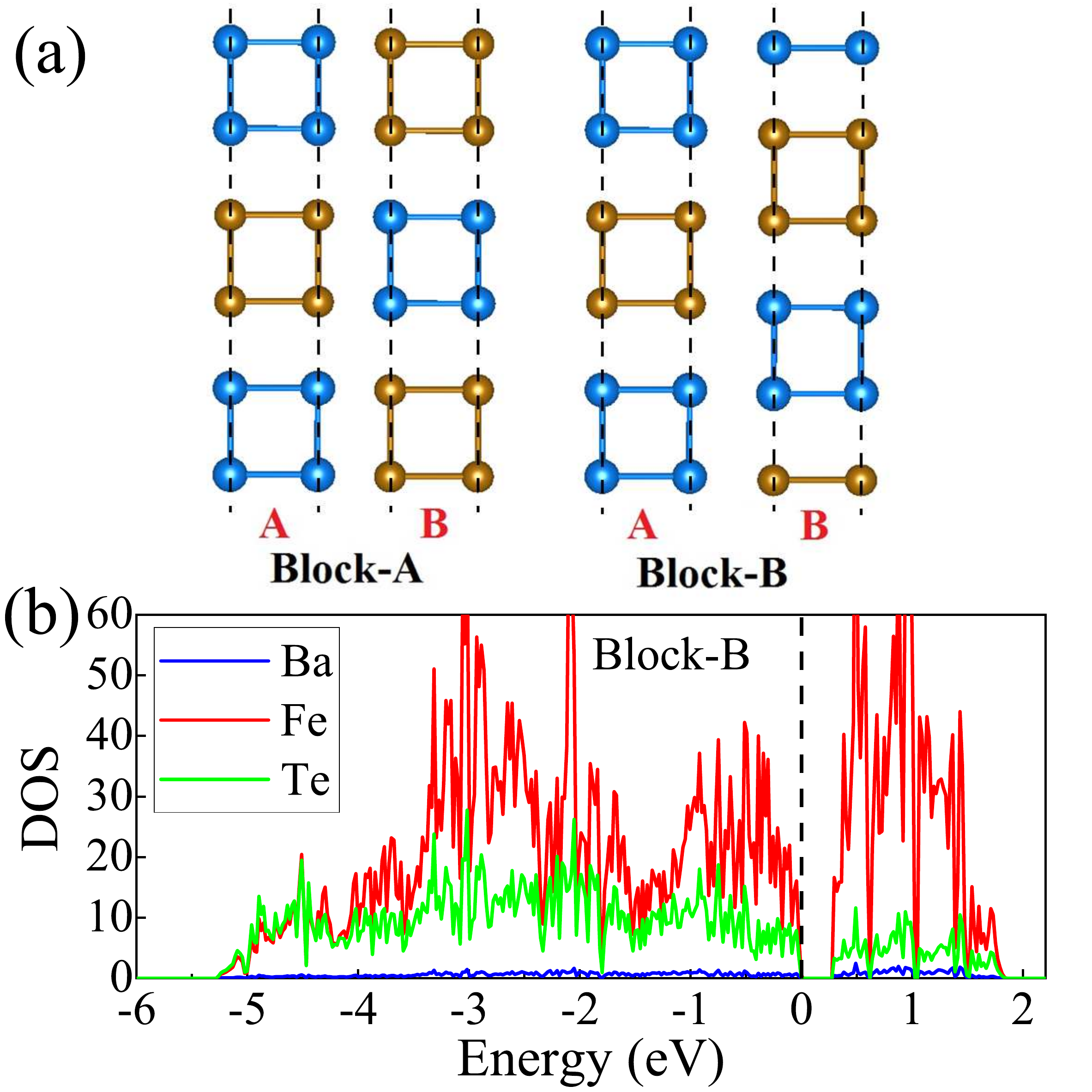}
\caption{(a) Sketch of Block-A and Block-B spin patterns studied here. Spin up and down are represented by blue and brown circles, respectively. The A and B red labels are different ladders located in different layers. The figure was reproduced from our previous publication~\cite{Zhang:prb18}. (b) The density-of-states near the Fermi level based on the Block-B states ($\pi$, $\pi$, $0$) for BaFe$_2$Te$_3$. Blue: Ba; Red: Fe; Green: Te.}
\label{Fig3}
\end{figure}

Previous studies of iron-based superconductors suggest that in the parent compound they all ordered magnetically.
Then, our next task is to understand the magnetic ground state of the here predicted new Te-based ladder. For this purpose,
various possible (in-ladder) magnetic arrangements were tried (see Fig.~S1 in SM~\cite{Supplemental}). Considering previous neutron-based studies for BaFe$_2$Se$_3$~\cite{Caron:Prb12}, the two possible Block AFM orders that were tried here are shown in Fig.~\ref{Fig3}(a). Our main results for BaFe$_2$Te$_3$ are summarized in Table~\ref{Table3}.

Under ambient conditions, our results indicate that the Block-B AFM order is the most stable ground-state magnetic order among all the candidates considered here. For the Block-B AFM state, the calculated local magnetic moment of Fe is $2.85$ $\mu_{\rm B}$/Fe, quite close to the value observed experimentally and theoretically for BaFe$_2$Se$_3$~\cite{Caron:Prb12,Zhang:prb18}. The calculated energy gap of the Block-B AFM order is about $0.32$~eV, which is slightly smaller than the calculated value of BaFe$_2$Se$_3$~\cite{Dong:PRL14,Zhang:prb18}. Moreover, BaFe$_2$Te$_3$ could become metallic under high pressure after considering previous DFT calculations and experiments for related ladders~\cite{Zhang:prb17,Zhang:prb18,Takahashi:Nm,Ying:prb17}. An structural phase transition in the case of BaFe$_2$Se$_3$ was found under pressure, with the tilting ladders becaming non-tilting, corresponding to a transition from the $Pbnm$ symmetry to an ideal $Cmcm$ symmetry~\cite{Svitlyk:JPCM,Zhang:prb18}. Hence, in principle BaFe$_2$Te$_3$ should also display an structural phase transition under pressure although this aspect
requires further detailed calculations beyond the scope of this publication.

\begin{table} [H]
\centering\caption{The optimized lattice constants ({\AA}), local magnetic moments (in $\mu_{\rm B}$/Fe units) within the default PAW sphere,
and band gaps (eV) for the various magnetic configurations considered, as well as the energy differences (meV/Fe) with respect to the Block-B configuration taken as the reference of energy. All the magnetic states discussed here were fully optimized starting from the $Pbnm$ structure.}
\begin{tabular*}{0.48\textwidth}{@{\extracolsep{\fill}}llllc}
\hline
\hline
  & $a$/$b$/$c$ & $M$ & Gap  & Energy \\
\hline
NM       & 9.177/12.200/5.658  & 0    & 0    & 385.2 \\
FM       & 9.852/13.397/5.532  & 2.90 & 0    & 125.4  \\
CX       & 9.729/13.131/5.652  & 2.63 & 0.05 & 56.6    \\
CY       & 9.894/13.178/5.590  & 2.79 & 0.23 & 51.5     \\
G        & 9.840/12.852/5.643  & 2.46 & 0.06 & 140.3   \\
Block-A  & 9.806/13.424/5.577  & 2.86 & 0.26 & 3.1     \\
Block-B  & 9.824/13.182/5.615  & 2.85 & 0.32 & 0        \\
\hline
\hline
\end{tabular*}
\label{Table3}
\end{table}

According to the calculated density of states (DOS) for the Block-B ($\pi$, $\pi$, $0$) AFM order [see Fig.~\ref{Fig3}(b)], the bands near the Fermi level are primarily contributed by Fe-$3d$ orbitals which are highly $hybridized$ with Te-$5p$ orbitals. The total bandwidth of the iron bands corresponding to the magnetic ground state of BaFe$_2$Te$_3$ ($\sim 7$ eV) is slightly smaller than the bandwidths of BaFe$_2$S$_3$ ($\sim 8$ eV)~\cite{Zhang:prb19} and BaFe$_2$Se$_3$ ($\sim 7.6$ eV)~\cite{bandwidthcontex}. This suggests that electronic correlation effects are enhanced in Te-based ladders because
the ratio $U/W$, with $U$ the local atomic Hubbard repulsion and $W$ the bandwidth, is effectively
enhanced. This conclusion agrees with our recent theoretical studies in $n=5.5$ iron Te-based ladders~\cite{Zhang:prb19}. Because many believe that the Block-type magnetic order of BaFe$_2$Se$_3$ is related to an orbital selective Mott state induced by electronic correlations~\cite{Caron:Prb12,osmp1,osmp2,osmp3}, it is reasonable to conclude that BaFe$_2$Te$_3$ could also display this interesting state as well. Of course, more powerful many-body techniques based on multiorbital Hubbard
models are required to confirm this OSMP hypothesis.

\subsection{D. Ferrielectricity}
\begin{figure}
\centering
\includegraphics[width=0.48\textwidth]{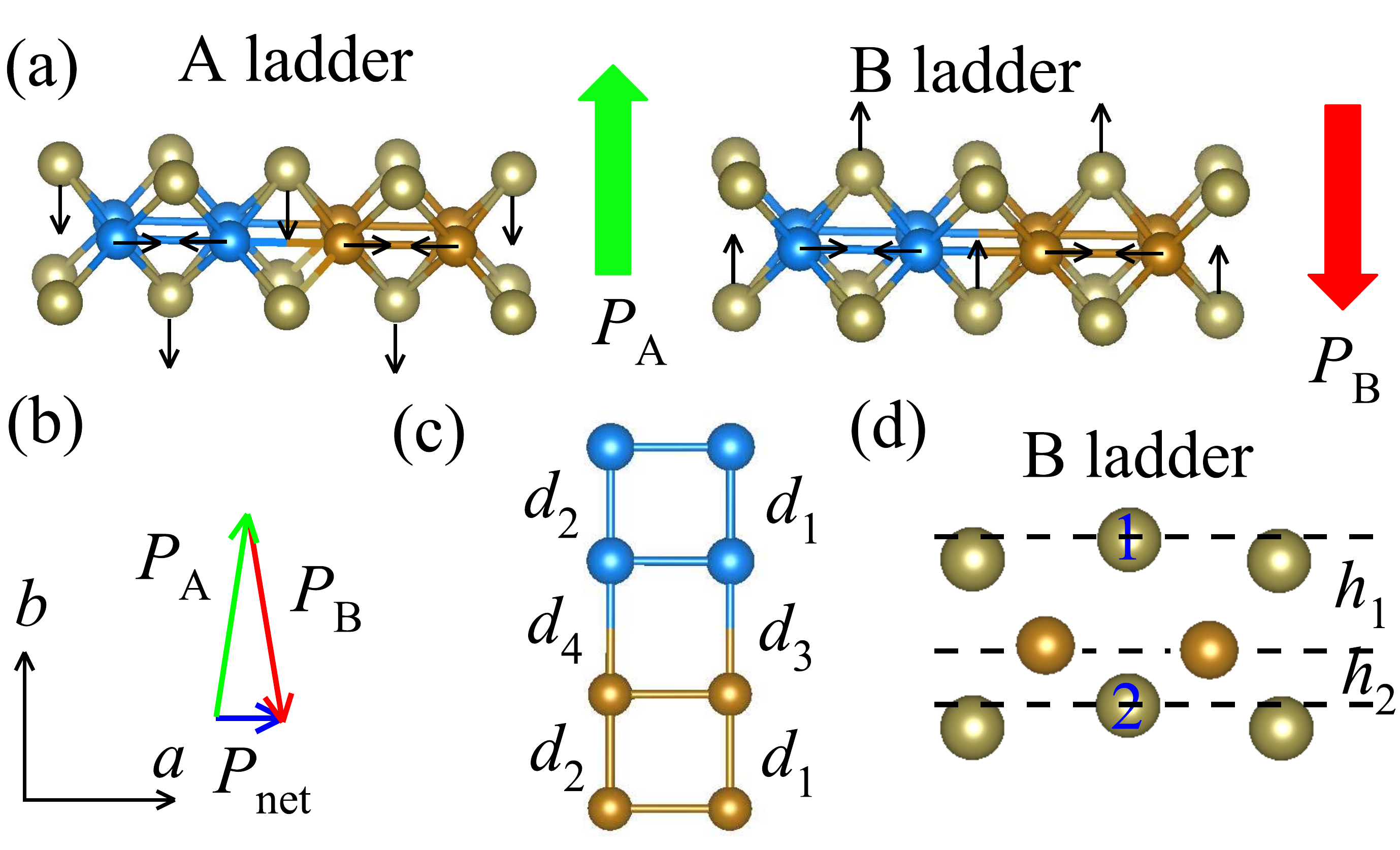}
\caption{(a) Sketch of the Fe-Te ladders A and B with Block AFM order. Partial Te atomic displacements induced by the exchange striction effect of the Block order of Fe. (b) Vector of polarizations of the different ladders A and B, as well as the net polarization. (c) Sketch of one ladder with the optimized Block-B AFM order of BaFe$_2$Te$_3$, showing the NN spin up-up (or down-down) and NN spin up-down Fe-Fe distances. (d) Sketch along Ladder B of the optimized Block-B AFM order of BaFe$_2$Te$_3$. The different heights of Te are marked $h_1$ and $h_2$, respectively.}
\label{Fig4}
\end{figure}

Due to magnetic-exchange striction effect, the Block spin configuration will break parity symmetry but will not break space-inversion symmetry in the iron ladder. However, as shown in Fig.~\ref{Fig4}(a), the displacements of the Te atoms would break space-inversion symmetry. This will induce local dipoles for each iron ladder since irons would move in the same direction perpendicular to the ladder's plane. Due to the phase difference between the A and B ladders in the Block-B AFM state of BaFe$_2$Te$_3$, the induced polarization of each ladder is in principle opposite~\cite{AAcontext}. However, there will be a remaining net polarization along the $c$-axis since the ladders A and B are slightly tilting, as displayed in Fig.~\ref{Fig4}~(b). This conclusion is also supported by group theory analysis~\cite{Groupcontext}.

In our fully optimized crystal structure of Block-B AFM order of BaFe$_2$Te$_3$, the nearest-neighbor (NN) distances of spin up-up [or down-down] Fe-Fe are $d_1$=$2.583$~\AA and $d_2$=$2.594$~\AA, and the NN distances of spin up-down Fe-Fe are $d_3$=$3.031$~\AA and $d_2$=$3.020$~\AA, as displayed in Fig.~\ref{Fig4}~(c). Meanwhile, our DFT results indicate that the heights of Te(1) and Te(2) are different: $1.987$~\AA and $1.454$~\AA, respectively, which is in agreement with our symmetry and group analysis mentioned in the previous paragraph. Those numerical results show that the ferroelectric dipole of BaFe$_2$Te$_3$ would be larger than the value of BaFe$_2$Se$_3$ by comparing the height difference of different chalcogens (for Te: $\Delta$$h$$\sim0.53$~\AA ~while for Se: $\Delta$$h$$\sim0.22$~\AA ~\cite{Dong:PRL14}). The calculated $P$ of BaFe$_2$Te$_3$ is about $0.31$ $\mu$C/cm$^2$, which is larger than the value for BaFe$_2$Se$_3$ ($0.19$ $\mu$C/cm$^2$)~\cite{Dong:PRL14}. The ferrielectric polarization is directly proportional to the effective ionic charge and relative displacement from paraelectric state, and inversely proportional to volume. Although the difference in distance of the heights of Te is more than twice that of the heights of Se, the value of the polarization of BaFe$_2$Te$_3$ is not twice as large as that of BaFe$_2$Se$_3$. This is because we also have to consider that the effective ionic charge within chalcogenides decreases from Se to Te~\cite{Weakelectronegativitycontext}, while the volume of BaFe$_2$Se$_3$ is smaller than Ba$_2$Fe$_2$Te$_3$.
%
More importantly, because the ladders for each layer are tilted in different directions, the corresponding induced ferroelectric dipoles would correspond to a non-collinear order as shown in Fig.~\ref{Fig4}(b). Its noncollinearities may be easily modulated by external electric fields. If we only considering the exchange striction without magnetic order, the space group of our fully relaxed FiE structure is $Pnm2_1$ (No.31) which is consistent with recent neutron experiments for BaFe$_2$Se$_3$~\cite{Aoyama:prb19}.

\section{IV. Additional discussion}
\begin{figure}
\centering
\includegraphics[width=0.48\textwidth]{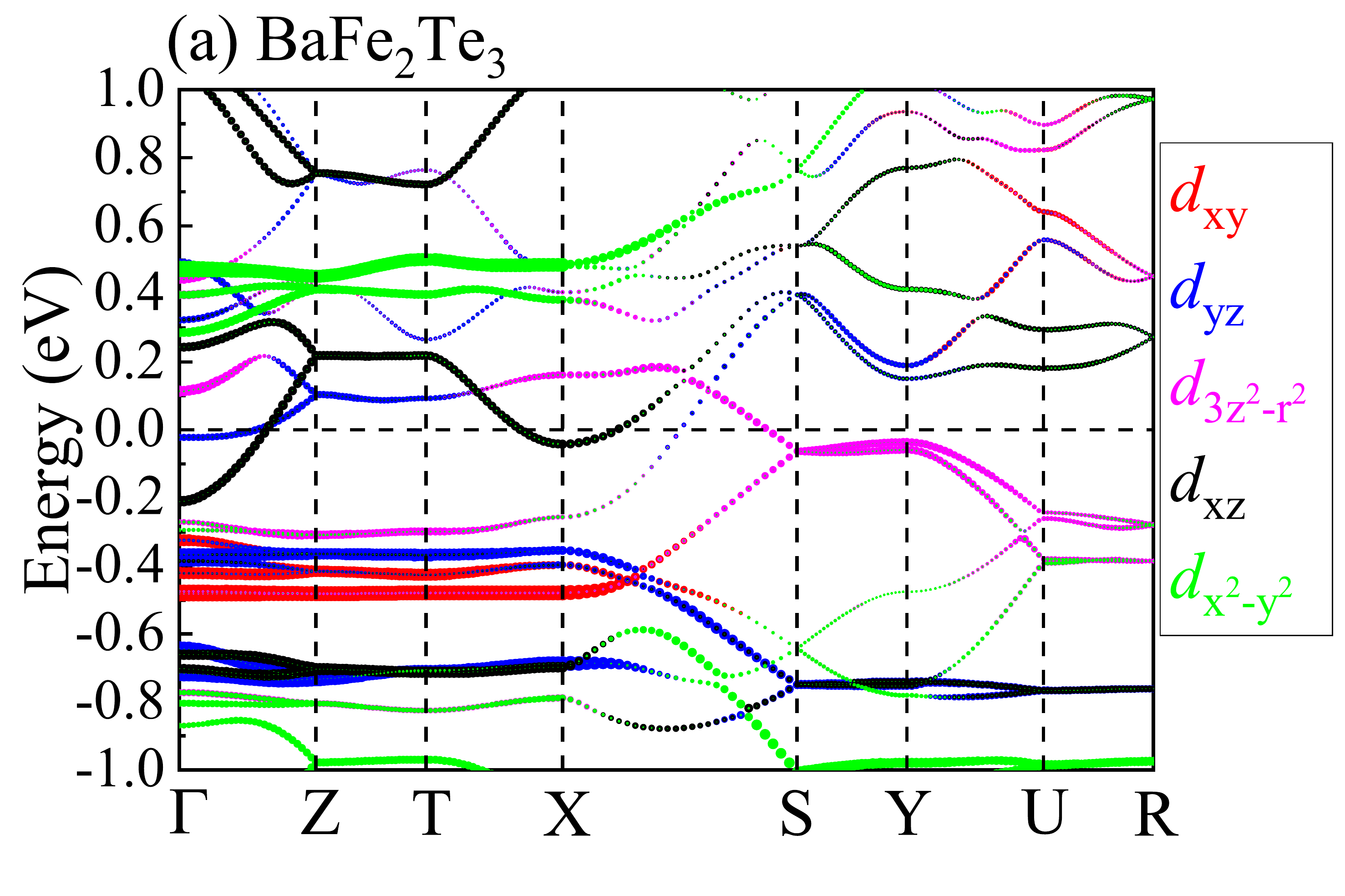}
\includegraphics[width=0.48\textwidth]{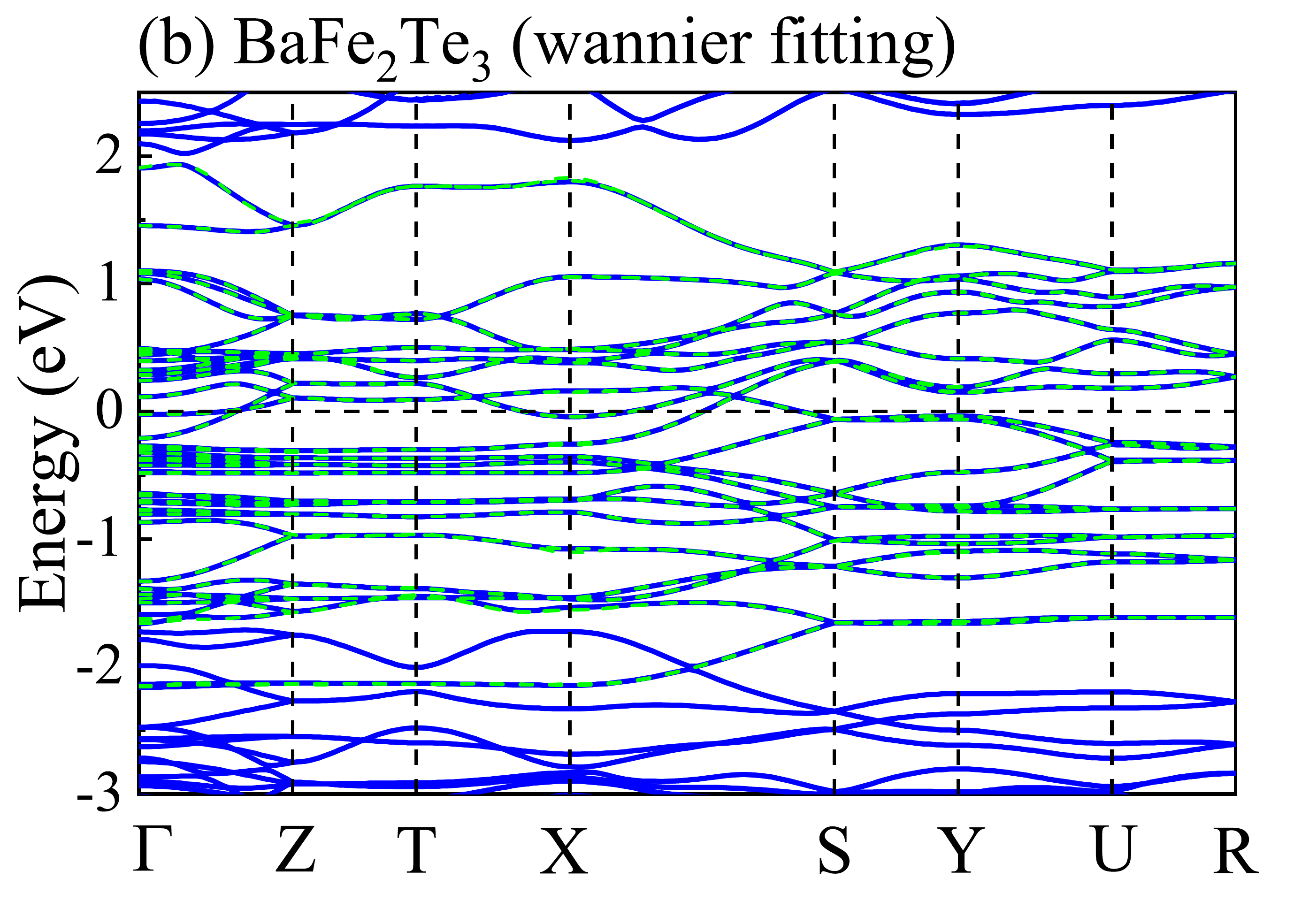}
\caption{ (a) Projected band structures of BaFe$_2$Te$_3$ (electronic density $n = 6$) for the non-magnetic (NM) state. The Fermi level is shown with dashed lines. The weight of each iron orbital is represented by the size of the circle. (b) The original band dispersion is shown by blue solid, while the Wannier interpolated band dispersion is shown using green dashed curves for BaFe$_2$Te$_3$. The coordinates of the high symmetry points in bulk BZ are given by: $\Gamma$ = (0, 0, 0), Z = (0, 0, 0.5), T = (-0.5, 0, 0.5), X = (-0.5 0 0), S = (-0.5, 0.5, 0), Y = (0, 0,5, 0), U = (0, 0.5, 0.5), R = (-0.5, 0.5, 0.5). }
\label{Fig5}
\end{figure}

\begin{figure}
\centering
\includegraphics[width=0.48\textwidth]{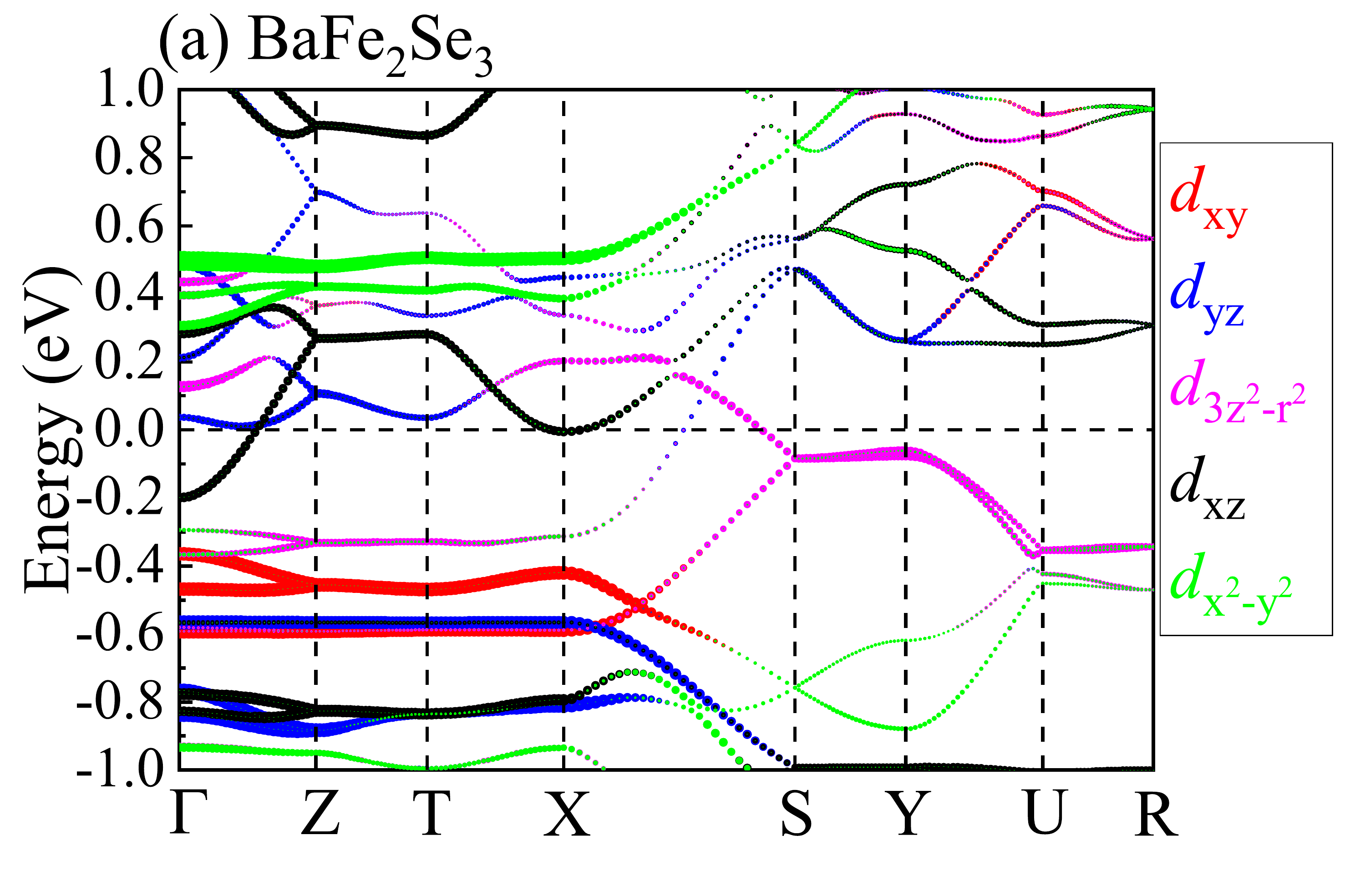}
\includegraphics[width=0.48\textwidth]{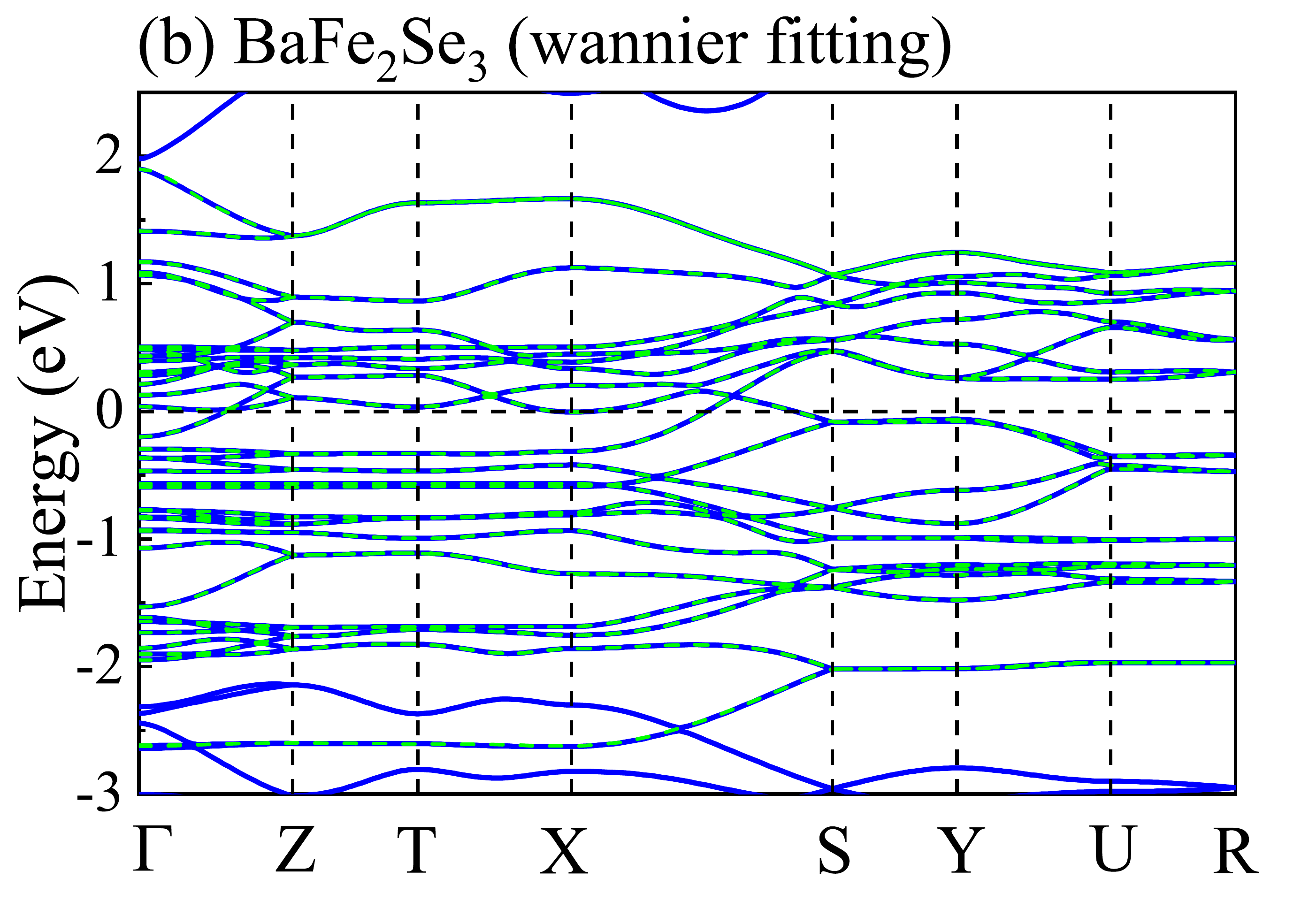}
\caption{ (a) Projected band structures of BaFe$_2$Se$_3$ (electronic density $n = 6$) for the non-magnetic (NM) state. The Fermi level is shown with dashed lines. The weight of each iron orbital is represented by the size of the circle. (b) The original band dispersion is shown by blue solid, while the Wannier interpolated band dispersion is shown using green dashed curves for BaFe$_2$Se$_3$. The coordinates of the high symmetry points in bulk BZ are given by: $\Gamma$ = (0, 0, 0), Z = (0, 0, 0.5), T = (-0.5, 0, 0.5), X = (-0.5 0 0), S = (-0.5, 0.5, 0), Y = (0, 0,5, 0), U = (0, 0.5, 0.5), R = (-0.5, 0.5, 0.5).}
\label{Fig6}
\end{figure}

In Fig.~\ref{Fig5}~(a), we present the projected band structure of the fully optimized non-magnetic states for the five iron $3d$ orbitals corresponding to BaFe$_2$Te$_3$. For better understanding, we changed the lattice vectors of the $Pbnm$ space group to the conventional $Pnma$ space group where the b-axis is along the ladders direction, the c-axis is perpendicular to the ladders but still in the iron layer, and the a-axis is perpendicular to the iron layer. It is clearly shown that the band structure is more dispersive along the ladder direction ($X-S$ path) than other directions, which indicates the quasi one-dimensional ladder behavior along the $k_y$ axis. Near the Fermi level, the bands are mainly contributed by $d_{xz}$, $d_{yz}$ and $d_{3z^2-r^2}$ (note that the cartesian axes correspond to the lattice axis of $Pnma$ symmetry~\cite{Caron:Prb}; i.e. the $x$ axis is $a$, $y$ axis is $b$ and $z$ axis is $c$ in Fig.~\ref{Fig1}~(b)). Based on the Wannier fitting results, the bandwidth of the five iron orbitals is about $4.1$~eV, as shown in Fig.~\ref{Fig5}~(b). Since the crystal constants of the fully optimized non-magnetic state are usually smaller than the experimental lattice constants in iron ladder systems, the ``real'' non-interacting bandwidth of the five iron orbitals would be smaller than this value. To better understand the electronic correlation of BaFe$_2$Te$_3$ and the electronic structural similarity with BaFe$_2$Se$_3$, we display the electronic structure of BaFe$_2$Se$_3$ for the fully optimized non-magnetic phase in Fig.~\ref{Fig6}. In BaFe$_2$Se$_3$, the band structure is quite similar to BaFe$_2$Te$_3$ because the Fermi level are also mainly contributed by $d_{xz}$, $d_{yz}$ and $d_{3z^2-r^2}$ orbitals. As shown in Fig.~\ref{Fig6}~(b), the bandwidth of the $123$-Se $n = 6$ ladder for the optimized non-magnetic structure is about $4.5$~eV. Because the electronic correlation strength is given by the ratio $U/W$, this analysis indicates that the degree of electronic correlation is enhanced in Te-ladders.

The vast majority of reported 2D iron superconductors~\cite{Stewart:Rmp,Johnston:Ap,Dagotto:Rmp,Dai:Rmp} display a similar structural lattice, magnetic ground state, and electronic structure. Then, it is reasonable that according to our DFT results, the physical and structural properties of our predicted BaFe$_2$Te$_3$ are very similar to those of BaFe$_2$Se$_3$. The BaFe$_2$S$_3$ has a CX-type AFM order at ambient conditions, and the superconducting transition temperature of BaFe$_2$S$_3$ is higher than BaFe$_2$Se$_3$ (the $2\times2$ Block-type AFM order). It suggests that BaFe$_2$Te$_3$ could also become superconducting under pressure although with lower transition temperature than BaFe$_2$S$_3$ because our 123-Te ladder is very similar to BaFe$_2$Se$_3$ which has a lower critical temperature than the 123-S ladder.

Since the pressure-induced superconducting phase domes of iron ladders were found near the AFM phase~\cite{Takahashi:Nm,Ying:prb17}, it is reasonable to assume that the driving force of superconductivity are the AFM spin fluctuations~\cite{Arita:prb,Zhang:prb18}. The Block AFM order of BaFe$_2$Se$_3$ was considered to possibly change to the stripe CX-AFM structure due to a structural transition under pressure~\cite{Ying:prb17,Zhang:prb18}. In this scenario, the magnetic fluctuations of BaFe$_2$Se$_3$ were considered to be induced by the competition between Block- and CX-type magnetic orders, while the magnetic fluctuations of BaFe$_2$S$_3$ were attributed entirely to the CX-type AFM order. Hence, it is reasonable to assume that the competition between Block and CX-type magnetic order would be the driving force for superconductivity in high-pressured BaFe$_2$Te$_3$ as well. It should also be noted that, in previous 2D iron superconductors, the $d_{xy}$ orbital was considered to be very important to understand the transition temperature~\cite{Hirschfeld:Rpp,Scalapino:Rmp}, corresponding to our $d_{yz}$ orbital lying in the iron ladder plane. This issue deserves to be studied more in-depth by powerful many-body techniques, beyond the capabilities of the DFT calculations.

\section{V. Conclusion}
In summary, the two-leg iron ladder compound BaFe$_2$Te$_3$, with the iron density $n=6$, was systematically studied by using first-principles calculations. The Block-B type AFM state is here predicted to be the most likely magnetic ground state. The decreased bandwidth of the iron $3d$ bands from S to Te indicates an enhanced degree of electronic correlation. Considering the exotic Block order and strong correlation in this $n=6$ iron Te ladder, the phenomenon of orbital-selective Mott phase is expected. In addition, using the symmetry analysis and DFT calculations, the presence of noncollinear ferrieletricity in BaFe$_2$Te$_3$ is here predicted as induced by the magnetic exchange striction effects of the Block order. Moreover, considering the magnetic state similarity and electronic structure with other iron ladders, BaFe$_2$Te$_3$ may become superconducting under higher pressure. Our overarching conclusion is that the $n=6$ iron Te ladder is worth to be studied by
theoretical and experimental procedures because using Te could lead to interesting results, such as exotic magnetic states, an orbital selective Mott phase, noncollinear ferrieletricity, as well as superconductivity under high pressure.

\section{Acknowledgments}
E.D. and A.M. were supported by the U.S. Department of Energy (DOE), Office of Science, Basic Energy Sciences (BES), Materials Sciences and Engineering Division. S.D., Y.Z., and L.F.L. were supported by the National Natural Science Foundation of China (Grant Nos. 11834002 and 11674055). L.F.L. and Y.Z. were supported by the China Scholarship Council. Y.Z. was also supported by the Scientific Research Foundation of Graduate School of Southeast University.  Most calculations were carried out at the Advanced Computing Facility (ACF) of the University of Tennessee Knoxville (UTK).


\begin{references}
\bibitem{Stewart:Rmp} G. R. Stewart, \href{https://doi.org/10.1103/RevModPhys.83.1589}{Rev. Mod. Phys. \textbf{83}, 1589 (2011).}
\bibitem{Johnston:Ap} D. C. Johnston, \href{https://doi.org/10.1080/00018732.2010.513480}{Adv. Phys. \textbf{59}, 803 (2010).}
\bibitem{Dagotto:Rmp} E. Dagotto, \href{https://doi.org/10.1103/RevModPhys.85.849}{Rev. Mod. Phys. \textbf{85}, 849 (2013).}
\bibitem{Dai:Rmp} P. C. Dai, \href{https://doi.org/10.1103/RevModPhys.87.855}{Rev. Mod. Phys. \textbf{87}, 855 (2015).}
\bibitem{Li:Np}W. Li, H. Ding, P. Deng, K. Chang, C. L. Song, K. He, L. L. Wang, X. C. Ma, J. P. Hu, X. Chen, and Q. K. Xue, \href{https://doi.org/10.1038/nphys2155}{Nat. Phys. \textbf{8}, 126 (2012).}
\bibitem{Zhang:RRL} Y. Zhang, H. M. Zhang, Y. K. Weng, L. F. Lin, X. Y. Yao, and S. Dong, \href{https://doi.org/10.1002/pssr.201600279}{Phys. Status Solidi RRL \textbf{10}, 757 (2016).}
\bibitem{Dai:Np} P. C. Dai, J. P. Hu, and E. Dagotto, \href{https://doi.org/10.1038/nphys2438}{Nat. Phys. \textbf{8}, 709 (2012).}
\bibitem{Mazin:np} I. I. Mazin and M. D. Johannes, \href{https://doi.org/10.1038/nphys1160}{Nat. Phys. \textbf{5}, 141 (2009).}
\bibitem{Takahashi:Nm} H. Takahashi, A. Sugimoto, Y. Nambu, T. Yamauchi, Y. Hirata, T. Kawakami, M. Avdeev, K. Matsubayashi, F. Du, C. Kawashima, H. Soeda, S. Nakano, Y. Uwatoko, Y. Ueda, T. J. Sato and K. Ohgushi, \href{https://doi.org/10.1038/nmat4351}{Nat. Mater. \textbf{14}, 1008 (2015).}
\bibitem{Ying:prb17} J.-J. Ying, H. C. Lei, C. Petrovic, Y.-M. Xiao and V.-V. Struzhkin, \href{https://doi.org/10.1103/PhysRevB.95.241109}{Phys. Rev. B \textbf{95}, 241109(R) (2017).}
\bibitem{Yamauchi:prl15} T. Yamauchi, Y. Hirata, Y. Ueda, and K. Ohgushi, \href{https://doi.org/10.1103/PhysRevLett.115.246402}{Phys. Rev. Lett. \textbf{115} 246402 (2015).}
\bibitem{Arita:prb} R. Arita, H. Ikeda, S. Sakai, and M.-To Suzuki, \href{https://doi.org/10.1103/PhysRevB.92.054515}{Phys. Rev. B \textbf{92}, 054515 (2015).}
\bibitem{Patel:prb16} N. D. Patel, A. Nocera, G. Alvarez, R. Arita, A. Moreo, and E. Dagotto, \href{https://doi.org/10.1103/PhysRevB.94.075119}{Phys. Rev. B \textbf{94}, 075119 (2016).}
\bibitem{Wang:prb16} M. Wang, M. Yi, S. J. Jin, H. C. Jiang,  Y. Song, H. Q. Luo, A. D. Christianson, C. de la Cruz, E. Bourret-Courchesne, D. X. Yao, D. H. Lee, and R. J. Birgeneau, \href{https://doi.org/10.1103/PhysRevB.94.041111}{Phys. Rev. B \textbf{94}, 041111(R) (2016).}
\bibitem{chi:prl} S. X. Chi, Y. Uwatoko, H. B. Cao, Y. Hirata, K. Hashizume, T. Aoyama, and K. Ohgushi, \href{https://doi.org/10.1103/PhysRevLett.117.047003}{Phys. Rev. Lett. \textbf{117}, 047003 (2016).}
\bibitem{Zhang:prb17} Y. Zhang, L. F. Lin, J. J. Zhang, E. Dagotto, and S. Dong, \href{https://doi.org/10.1103/PhysRevB.95.115154}{Phys. Rev. B \textbf{95}, 115154 (2017).}
\bibitem{Patel:prb17} N. D. Patel, A. Nocera, G. Alvarez, A. Moreo, and E. Dagotto, \href{https://doi.org/10.1103/PhysRevB.96.024520}{Phys. Rev. B \textbf{96}, 024520 (2017).}
\bibitem{Pizarro:prm} J. M. Pizarro and E. Bascones, \href{https://doi.org/10.1103/PhysRevMaterials.3.014801}{Phys. Rev. Mater. \textbf{3}, 014801 (2019).}
\bibitem{Zheng:prb18} L. Zheng, B. A. Frandsen, C. Wu, M. Yi, S. Wu, Q. Huang, E. Bourret-Courchesne, G. Simutis, R. Khasanov, D.-X. Yao, M. Wang, and R. J. Birgeneau, \href{https://doi.org/10.1103/PhysRevB.98.180402}{Phys. Rev. B \textbf{98}, 180402(R) (2018).}
\bibitem{Zhang:prb18} Y. Zhang, L. F. Lin, J. J. Zhang, E. Dagotto, and S. Dong, \href{https://doi.org/10.1103/PhysRevB.97.045119}{Phys. Rev. B \textbf{97}, 045119 (2018).}
\bibitem{Zhang:prb19} Y. Zhang, L. F. Lin, A. Moreo, S. Dong, and E. Dagotto, \href{https://doi.org/10.1103/PhysRevB.100.184419}{Phys. Rev. B \textbf{100}, 184419 (2019).}
\bibitem{Suzuki:prb} M. T. Suzuki, R. Arita, and H. Ikeda, \href{https://doi.org/10.1103/PhysRevB.92.085116}{Phys. Rev. B \textbf{92}, 085116 (2015).}
\bibitem{Materne:prb19} P. Materne, W. Bi, J. Zhao, M. Y. Hu, M. L. Amig\'o, S. Seiro, S. Aswartham,
B. B\"uchner, and E. E. Alp, \href{https://doi.org/10.1103/PhysRevB.99.020505}{Phys. Rev. B \textbf{99}, 020505(R) (2019).}
\bibitem{cu-ladder1} E. Dagotto, J. Riera, and D. Scalapino, \href{https://doi.org/10.1103/PhysRevB.45.5744}{Phys. Rev. B {\bf 45}, 5744(R) (1992).}
\bibitem{cu-ladder2} E. Dagotto and T. M. Rice, \href{https://doi.org/10.1126/science.271.5249.618}{Science {\bf 271}, 618 (1996)}. See also
E. Dagotto, \href{https://doi.org/10.1088/0034-4885/62/11/202}{Rep. Prog. Phys. {\bf 62}, 1525 (1999).}
\bibitem{cu-ladder3} M. Uehara, T. Nagata, J. Akimitsu, H. Takahashi, N. Mori, and K. Kinoshita, \href{https://doi.org/10.1143/JPSJ.65.2764}{J. Phys. Soc. Jpn. {\bf 65}, 2764 (1996).}
\bibitem{Caron:Prb} J. M. Caron, J. R. Neilson, D. C. Miller, A. Llobet, and T. M. McQueen, \href{https://doi.org/10.1103/PhysRevB.84.180409}{Phys. Rev. B \textbf{84}, 180409(R) (2011).}
\bibitem{Caron:Prb12} J. M. Caron, J. R. Neilson, D. C. Miller, K. Arpino, A.  Llobet, and T. M. McQueen, \href{https://doi.org/10.1103/PhysRevB.85.180405}{Phys. Rev. B \textbf{85}, 180405(R) (2012).}
\bibitem{Nambu:Prb} Y. Nambu, K. Ohgushi, S. Suzuki, F. Du, M. Avdeev, Y. Uwatoko, K. Munakata, H. Fukazawa, S. X. Chi, Y. Ueda, and T. J. Sato, \href{https://doi.org/10.1103/PhysRevB.85.064413}{Phys. Rev. B \textbf{85}, 064413 (2012).}
\bibitem{mourigal:prl15} M. Mourigal, Shan Wu, M. B. Stone, J. R. Neilson, J. M. Caron, T. M. McQueen, and C. L. Broholm, \href{https://doi.org/10.1103/PhysRevLett.115.047401}{Phys. Rev. Lett. \textbf{115}, 047401 (2015).}
\bibitem{osmp1} J. Herbrych, N. Kaushal, A. Nocera, G. Alvarez, A. Moreo, and E. Dagotto,
\href{https://doi.org/10.1038/s41467-018-06181-6}{Nat. Comm. {\bf 9}, 3736 (2018).}
\bibitem{osmp2} N. D. Patel, A. Nocera, G. Alvarez, A. Moreo, S. Johnston, and E. Dagotto,
\href{https://doi.org/10.1038/s42005-019-0155-3}{Commun. Phys. {\bf 2}, 64 (2019).}
\bibitem{osmp3} J. Herbrych, J. Heverhagen, N. D. Patel, G. Alvarez, M. Daghofer, A. Moreo, and E. Dagotto, \href{https://doi.org/10.1103/PhysRevLett.123.027203}{Phys. Rev. Lett. {\bf 123}, 027203 (2019).}
\bibitem{Aoyama:prb19} T. Aoyama, S. Imaizumi, T. Togashi, Y. Sato, K. Hashizume, Y. Nambu, Y. Hirata, M. Matsubara, and K. Ohgushi,  \href{https://doi.org/10.1103/PhysRevB.99.241109}{Phys. Rev. B \textbf{99}, 241109(R) (2019).}
\bibitem{Dong:PRL14} S. Dong, J. M. Liu, and E. Dagotto, \href{https://doi.org/10.1103/PhysRevLett.113.187204}{Phys. Rev. Lett. \textbf{113}, 187204 (2014).}
\bibitem{Lin:PRL} L.-F.  Lin, Y. Zhang, A. Moreo, E. Dagotto, and S. Dong, \href{https://doi.org/10.1103/PhysRevLett.123.067601}{Phys. Rev. Lett. \textbf{123}, 067601 (2019).}
\bibitem{Yang:PRL} Y. Yang, J. \'I\~niguez, A.-J. Mao, and L. Bellaiche, \href{https://doi.org/10.1103/PhysRevLett.112.057202}{Phys. Rev. Lett. \textbf{112}, 057202 (2014).}
\bibitem{luo:prb13} Q. Luo, A. Nicholson, J. Rinc\'on, S. Liang, J. Riera, G. Alvarez, L. Wang, W. Ku, G. D. Samolyuk, A. Moreo, and E. Dagotto, \href{https://doi.org/10.1103/PhysRevB.87.024404}{Phys. Rev. B. \textbf{87}, 024404 (2013).}
\bibitem{Klepp:JOAC} K. O. Klepp, W. Sparlinek, and H. Boller, \href{https://doi.org/10.1016/0925-8388(95)02087-X}{J. Alloy. Compd. \textbf{238}, 1 (1996).}
\bibitem{Guan:prl} J. Guan, Z. Zhu, and D. Tom\'anek, \href{https://doi.org/10.1103/PhysRevLett.113.046804}{Phys. Rev. Lett. \textbf{113}, 046804 (2014).}
\bibitem{Guan:nl} J. Guan, D. Liu, Z. Zhu, and D. Tom\'anek, \href{https://doi.org/10.1021/acs.nanolett.6b00767}{Nano Lett. \textbf{16}, 3247 (2016).}
\bibitem{Tan:AM} W. C. Tan, Y. Cai, R. Ng, L. Huang, X. Feng, G. Zhang, Y.-W. Zhang, C. A. Nijhuis, X. Liu, and K.-W. Ang,  \href{https://doi.org/10.1002/adma.201700503}{Adv. Mater. \textbf{29}, 1700503 (2017).}
\bibitem{Kresse:Prb} G. Kresse and J. Hafner, \href{https://doi.org/10.1103/PhysRevB.47.558}{Phys. Rev. B \textbf{47}, 558 (1993).}
\bibitem{Kresse:Prb96} G.~Kresse and J.~Furthm\"{u}ller, \href{https://doi.org/10.1103/PhysRevB.54.1169}{Phys. Rev. B \textbf{54}, 11169 (1996).}
\bibitem{Blochl:Prb} P. E. Bl\"{o}chl, \href{https://doi.org/10.1103/PhysRevB.50.17953}{Phys. Rev. B \textbf{50}, 17953 (1994).}
\bibitem{Perdew:Prl} J. P. Perdew, K. Burke, and M. Ernzerhof, \href{https://doi.org/10.1103/PhysRevLett.77.3865}{Phys. Rev. Lett. \textbf{77}, 3865 (1996).}
\bibitem{Supplemental} For more results, see Supplemental Material at \href{http://link.aps.org/supplemental/10.1103/PhysRevB.xx/xxxxxx}{http://link.aps.org/supplemental/10.1103/PhysRevB.xx/xxxxxx.}
\bibitem{Chaput:prb} L. Chaput, A. Togo, I. Tanaka, and G. Hug, \href{https://doi.org/10.1103/PhysRevB.84.094302}{Phys. Rev. B \textbf{84}, 094302 (2011).}
\bibitem{Togo:sm} A. Togo, I. Tanaka, \href{https://doi.org/10.1016/j.scriptamat.2015.07.021}{Scr. Mater. \textbf{108}, 1 (2015).}
\bibitem{King-Smith:Prb} R. D. King-Smith and D. Vanderbilt, \href{https://doi.org/10.1103/PhysRevB.47.1651}{Phys. Rev. B \textbf{47}, 1651 (1993).}
\bibitem{Resta:Rmp} R. Resta,  \href{https://doi.org/10.1103/RevModPhys.66.899}{Rev. Mod. Phys. \textbf{66}, 899 (1994).}
\bibitem{Mostofi:cpc} A. A. Mostofi, J. R. Yates, Y. S. Lee, I. Souza, D. Vanderbilt, and N. Marzari, Comput. \href{https://doi.org/10.1016/j.cpc.2007.11.016}{Phys. Commun. \textbf{178}, 685 (2007).}
\bibitem{Svitlyk:JPCM} V. Svitlyk, D. Chernyshov, E. Pomjakushina, A. Krzton-Maziopa, K. Conder, V. Pomjakushin, R. P{\"o}ttgen. and V. Dmitriev, \href{https://doi.org/10.1088/0953-8984/25/31/315403}{J. Phys. Condens. Matter \textbf{25}, 315403 (2013).}
\bibitem{Orobengoa:jac} D. Orobengoa, C. Capillas, M. I. Aroyo, and J. M. Perez-Mato, \href{https://doi.org/10.1107/S0021889809028064}{J. Appl. Crystallogr. \textbf{42}, 820 (2009).}
\bibitem{Perez-Mato:aca} J. Perez-Mato, D. Orobengoa, and M. I. Aroyo, \href{https://doi.org/10.1107/S0108767310016247}{Acta Crystallogr. A \textbf{66}, 558 (2010).}
\bibitem{Mouhat:PRB} F. Mouhat and F.-X. Coudert, \href{https://doi.org/10.1103/PhysRevB.90.224104}{Phys. Rev. B \textbf{90}, 224104 (2014).}
\bibitem{bandwidthcontex} Based on the previously optimized magnetic ground state of BaFe$_2$Se$_3$~\cite{Zhang:prb18}, we calculated the DOS for the Block-B AFM ($\pi$, $\pi$, $0$) magnetic order, as shown in Fig.~S2 of the SM~\cite{Supplemental}.
\bibitem{AAcontext} Although there is a $\pi$-phase shift in each AA/BB ladder plane for the Block-B AFM ($\pi$, $\pi$, $0$) order, the movements of Te atoms, induced by the Block order, are in the same direction as shown in Fig.~S3 of the SM~\cite{Supplemental}. Hence, the $\pi$-phase shift will not change the direction of $P$.
\bibitem{Groupcontext} Using the crystal lattice plus the Block-B AFM order, the magnetic space group became $Cc$ (No.9) after fully relaxing the structure. The corresponding point group of the fully relaxed Block-B AFM state is $m$. A nonzero polarization $P$ is
    allowed in this polar point group.
\bibitem{Weakelectronegativitycontext} Here, we present the electronic density difference between BaFe$_2$Se$_3$ and BaFe$_2$Te$_3$ in the SM~\cite{Supplemental}. It is clearly shown that the Se anions with bright red spheres attracts more electrons than the Te anions, while the electronic density difference of the iron ladder plane indicated that the Fe cations lost more electrons in the BaFe$_2$Se$_3$ system. In this scenario, it is suggested that the electronegativity of Te is weaker than Se.
\bibitem{Hirschfeld:Rpp} P. J. Hirschfeld, M. M. Korshunov and I. I. Mazin, \href{https://doi.org/10.1088/0034-4885/74/12/124508}{Rep. Prog. Phys. \textbf{74}, 124508 (2011).}
\bibitem{Scalapino:Rmp} D. J. Scalapino, \href{https://doi.org/10.1103/RevModPhys.84.1383}{Rev. Mod. Phys. \textbf{84}, 1383 (2012).}

\end{references}
\end{document}